\pdfoutput=1
\documentclass[superscriptaddress,preprint]{revtex4-1}

\usepackage{graphicx}

\newcommand{\uso}{\affiliation{Chemistry, University of Southampton, Highfield, Southampton SO17 1BJ, UK}}%
\newcommand{\cfeldesy}{\affiliation{Center for Free-Electron Laser Science, DESY, Notkestrasse 85,
      22607 Hamburg, Germany}}%
\newcommand{\clf}{\affiliation{Central Laser Facility, STFC Rutherford Appleton Laboratory, Didcot, Oxfordshire OX11 0QX, UK}}%

\begin{document}

\title{Resonant multiphoton ionisation probe of the photodissociation dynamics of ammonia}

\author{Adam D. Smith}\uso
\author{Hannah M. Watts}\uso
\author{Edward Jager}\uso
\author{Daniel A. Horke}\cfeldesy
\author{Emma Springate}\clf
\author{Oliver Alexander}\clf
\author{Cephise Cacho}\clf
\author{Richard T. Chapman}\clf
\author{Russell S. Minns}\uso\email[Email: ]{r.s.minns@soton.ac.uk}

\begin{abstract}
  The dissociation dynamics of the $\tilde{A}$-state of ammonia have been studied using a resonant multiphoton ionisation probe in a photoelectron spectroscopy experiment. The use of a resonant intermediate in the multiphoton ionisation process changes the ionisation propensity, allowing access to different ion states when compared with equivalent single photon ionisation experiments. Ionisation through the $E'$ $^1$A$_1'$ Rydberg intermediate means we maintain overlap with the ion state for an extended period allowing us to monitor the excited state population for several hundred femtoseconds. The vibrational states in the photoelectron spectrum show two distinct timescales, 200 fs and 320 fs, that we assign to the non-adiabatic and adiabatic dissociation processes respectively. The different timescales derive from differences in the wavepacket trajectories for the two dissociation pathways that resonantly excite different vibrational states in the intermediate Rydberg state. The timescales are similar to those obtained from time resolved ion kinetic energy release measurements, suggesting we can measure the different trajectories taken out to the region of conical intersection.
\end{abstract}
\maketitle

\section{Introduction}
The UV photolysis of ammonia is a benchmark system for the study of non-adiabatic dynamics. The dissociation dynamics in the $\tilde{A}$-state involve transitions occurring at a conical intersection. Competing non-adiabatic and adiabatic dissociation processes lead to hydrogen abstraction in conjunction with ground or excited state NH$_2$ molecular fragments respectively.\cite{Stavros2012,Truhlar2009,Wells2009,Crim2008,Truhlar2007,Crim2006,Leone2000,Mordaunt1998,Biesner89_2,Ashfold85,Ashfold86,Bach2003,BachVibrational2002,Dixon1996,Seideman1995,Ziegler1985,Rodriguez2014,Yu2014} The relative simplicity of the molecular structure therefore belies rather complex photochemistry which means this small molecular system is still actively investigated. Part of the reason for this continued interest is the lack of universal detection techniques that can provide clear information about the full reaction mechanism. To increase the observation window attainable with a phototelectron spectroscopy probe, we use a resonant multiphoton ionisation probe at 400 nm. By ionising through the $E'$ $^1$A$_1'$ Rydberg state we access a different set of ion vibrational states, when compared with direct ionisation of the $\tilde{A}$-state and maximise the time period over which ionisation can occur.

Much of what we know about the excited state potentials important to the photodissociation dynamics of the ammonia $\tilde{A}$-state, come from calculations of the potential energy surfaces.\cite{Worth2011,Truhlar2007,Dixon1996,bernardi_potential_1996,Truhlar2009} The theoretical studies have highlighted many of the structural and dynamical features important in the dissociation process. The electronic ground state of ammonia has an equilibrium pyramidal structure while the excited $\tilde{A}$-state has a planar equilibrium structure. This leads to a vibrational progression in the umbrella mode, $\nu_2$, which dominates the UV absorption spectrum. The electronic excitation process involves the promotion of a lone pair electron from the nitrogen atom into a Rydberg type orbital of 3\textit{s} character. Extending the N-H bond length leads to a small barrier which is large enough to make the lowest two vibrational states of the dominant $\nu_2$ mode bound. As the H-NH$_2$ bond is extended further the orbital increases in $\sigma^*$ character leading to dissociation. At extended N-H bond lengths the ground and excited state potentials cross at a conical intersection where, depending on the energy available and how the molecule traverses this region, the molecule can dissociate into ground or excited state molecular fragments. In full dimensional calculations of the potential energy surface, the conical intersection is only seen at planar geometries,\cite{Truhlar2007} while the threshold for excited state product formation is 6.02 eV, which lies significantly above the barrier to dissociation (5.94 eV) seen at shorter N-H bond lengths.

Initial spectroscopic measurements based on absorption suggested lifetimes in the $\tilde{A}$-state were are on the order of 35 fs.\cite{Ziegler1985,Ashfold85,Ashfold86} While subsequent measurement and analysis of the energy partitioning between the internal molecular motions of the NH$_2$ fragment and the kinetic energy released\cite{DONNELLY1979,Biesner89_2,Biesner1988,Rodriguez2014} demonstrate that tunneling leads to vibrationaly cold molecular fragments and that the vibrational energy in the NH$_2$ fragment increases with the photon energy.\cite{Biesner89_2} Once the energy is sufficient to drive formation of the excited state NH$_2$ fragment this process is observed with the relative proportion of excited state fragment shown to generally increase with increasing energy.\cite{Biesner89_2} The dynamics can therefore be crudely split into three regions: below the barrier to dissociation, tunneling controls the dissociation timescale resulting in very low levels of vibrational excitation in the ground state NH$_2$ fragment produced; As higher vibrational states are excited, over the barrier dissociation becomes possible and a shorter dissociation time is expected and observed. At energies above the barrier but below the adiabatic dissociation threshold ground state dissociation products are formed. As the excitation energy is increased above 6.02 eV, excited state fragments can also be formed and the two channels effectively compete with the dynamics at the region of the conical intersection controlling which dominates.

More recently time resolved photoelectron spectroscopy and time resolved kinetic energy release measurements have been applied to this problem. The kinetic energy release measurements have used resonant ionisation of the hydrogen fragment to highlighted the different lifetimes associated with different levels of internal energy in the molecular fragments produced.\cite{Wells2009,Stavros2012,chatterley2013,Yu2014} The different lifetimes have been correlated with various decay pathways, with some invoking complex dynamics around the conical intersection as a possible intermediate where population can be trapped for several hundred femtoseconds before dissociation occurs.\cite{chatterley2013} The effectiveness of photoelectron spectroscopy has by comparison been limited by the often restricted observation window afforded by the available UV probes. Photoelectron spectroscopy is in principle a universal detection technique as all states can be ionised. The practical limitations derive from the changing ionisation potential as a function of molecular geometry such that with conventional laser sources the overlap of the excited state wavepacket with the molecular ion states is often rather short lived. Combined with this, it is often desireable to reduce the background contribution to the retrieved photoelectron spectrum such that energy of the probe photons is often lower than the maximum available. In ammonia the use of a lower energy stops the probe from exciting the $\tilde{A}$-state, which would make measuring the early time dynamics difficult due to significant probe pump contributions at early times. This restricts the probe wavelength to being >215 nm. Recent experiments have used a 240 nm probe which restricted the observation window to the first 75 fs following excitation.\cite{Stavros2012} The ion kinetic energy release measurements however suggest the lifetime associated with dissociation can be up to 400 fs for some dissociation processes.\cite{chatterley2013} This means there is a large component of the excited state dynamics not being measured by either technique. Increasing the energy of the probe to 200 nm is unlikely to increase the observation window extensively and when combined with the issues surrounding the analysis of the early time dynamics is unlikely to provide any extra information from a probe at 240 nm. The restriction on the observation window can potentially be removed with a very high energy probe such that all reactants and products can be measured, however these experiments are extremely complex.\cite{sekikawa1,sekikawa2,leoneXUV} A potential route around this is to use multiphoton ionisation of the excited state population while maintaining an almost background free signal. The information content can often be maximised if one of the absorption processes is resonant with an intermediate Rydberg state. The time dependence of the resonance conditions provides a rich data set to analyse the time dependence of the signal. In the experiments described below, we use a multiphoton probe at 400 nm to increase the observation window available to photoelectron spectroscopy measurements and can monitor excited state population out to several hundred fs. The absorption of the first probe photon populates the $E'$ $^1$A$_1'$ Rydberg state that allows for an extended view of the excited state potential surface. By ionising through the Rydberg state we also access a different set of final ion states, allowing access to the lowest vibrational levels of the molecular ion. The use of a resonant multiphoton probe therefore maximises our observation window through changes in the Franck-Condon (FC) factors involved in the ionisation process maintaining overlap with the accessible ion states for extended period. The measurements show none of the complications associated with high energy single photon probes and provide vibrational state dependent lifetimes that allow us to separate the adiabatic and non-adiabatic pathways to dissociation.

\section{Experimental}
An amplified femtosecond laser system (Red Dragon, KM Labs) generates 30 fs pulses of 800 nm light, with a pulse energy of up to 10 mJ and at a repetition rate of 1 KHz. The pump and probe arms of the experiment are separated before compression, allowing for independent control of the pulse energy and compression. The pump pulse is produced via fourth harmonic generation (FHG) of the fundamental (800 nm) beam, generating photons at around 200 nm. The 200 nm beam is produced using standard non-linear optics with sequential second, third and fourth harmonic generation in BBO. The energy per pulse generated at the fourth harmonic is controlled through the use of a waveplate in the third harmonic generation arm which controls the efficiency of this mixing process. The pulse energy is kept to $\sim$2 $\mu$J, which maximised the pump probe signal while maintaining a low two pump-photon ionisation background signal. The pump beam is focused to the centre of the interaction region of the VMI spectrometer, with a 1 m focal length mirror where it intersects the molecular beam. To generate the probe, 1 mJ of the 800 nm beam is frequency doubled to produce approximately 5 $\mu$J at 400 nm. The absolute pulse energy is controlled using a half waveplate before the doubling crystal to control the phase matching and obtain the desired intensity at the interaction region. The probe beam is reflection focused, with a 1 m torroidal mirror, to the centre of the interaction region of the VMI spectrometer where it crosses the pump at a small angle (3$^\circ$). The crossing angle was minimised to obtain greater overlap of the beams leading to increased signal to noise ratios and shorter collection times. The pump-probe delay is controlled using a motorised translation stage and spatial overlap of the pump and probe beams was checked using a LuAG:Ce phosphorescent crystal that can be placed at the centre of the interaction region of the VMI. The phosphorescence from the crystal was imaged using a electron multiplying charge-coupled device (EM-CCD). The pump and probe beams are both linearly polarised in the plane of the VMI detector, perpendicular to the time of flight axis.

The molecular beam is generated through the expansion of 5 \% NH$_3$ in He at  1 bar through a 1 kHz pulsed nozzle (Amsterdam cantilever\cite{Janssen2009}) with a 100 $\mu$m aperture. The resulting expansion passes through a 1 mm skimmer and enters the interaction region of the spectrometer through a hole in the centre of the repeller plate of the VMI. The expansion pressure is kept intentionally low to avoid the formation of dimers and larger molecular clusters. The VMI spectrometer is of the standard 3 plate Eppink and Parker design\cite{Eppink1997}. The detector consist of a 70 mm diameter microchannel plate (MCP) and a phosphor screen which is imaged using a sCMOS camera (PCO Gold Edge 5.5). Tuning of the BBO crystals in the FHG set-up allows for some tuning of the central wavelength of the pump. Through this process the pump was tuned to the $\nu_2 ^4$ level of the NH$_3$ $\tilde{A}$-state at 200.8 nm. This process initiates the dissociation process which is subsequently probed using a time delayed 400 nm pulse which drives the multiphoton ionisation process. The absorption of two or more photons leads to ionisation into both the ground and first excited ion state of ammonia. The excitation scheme used is outlined in figure \ref{PES-ass} A. The raw images obtained are background subtracted before being deconvoluted using a polar onion peeling method,\cite{POP} providing both the photoelectron spectrum and the photoelectron angular distributions. Although the photoelectron angular distributions are obtained, they show no time dependence and as such are not discussed in the results section. The background subtraction involves subtracting the pump-probe images for very negative delays (i.e. probe arriving before the pump) corresponding to pump only and probe only signals. Calibration of the retrieved spectrum is done using XUV ionisation of noble gases using a high harmonic generation source.

\section{Results and Discussion}

\begin{figure} \centering
\includegraphics[width=0.7\columnwidth]{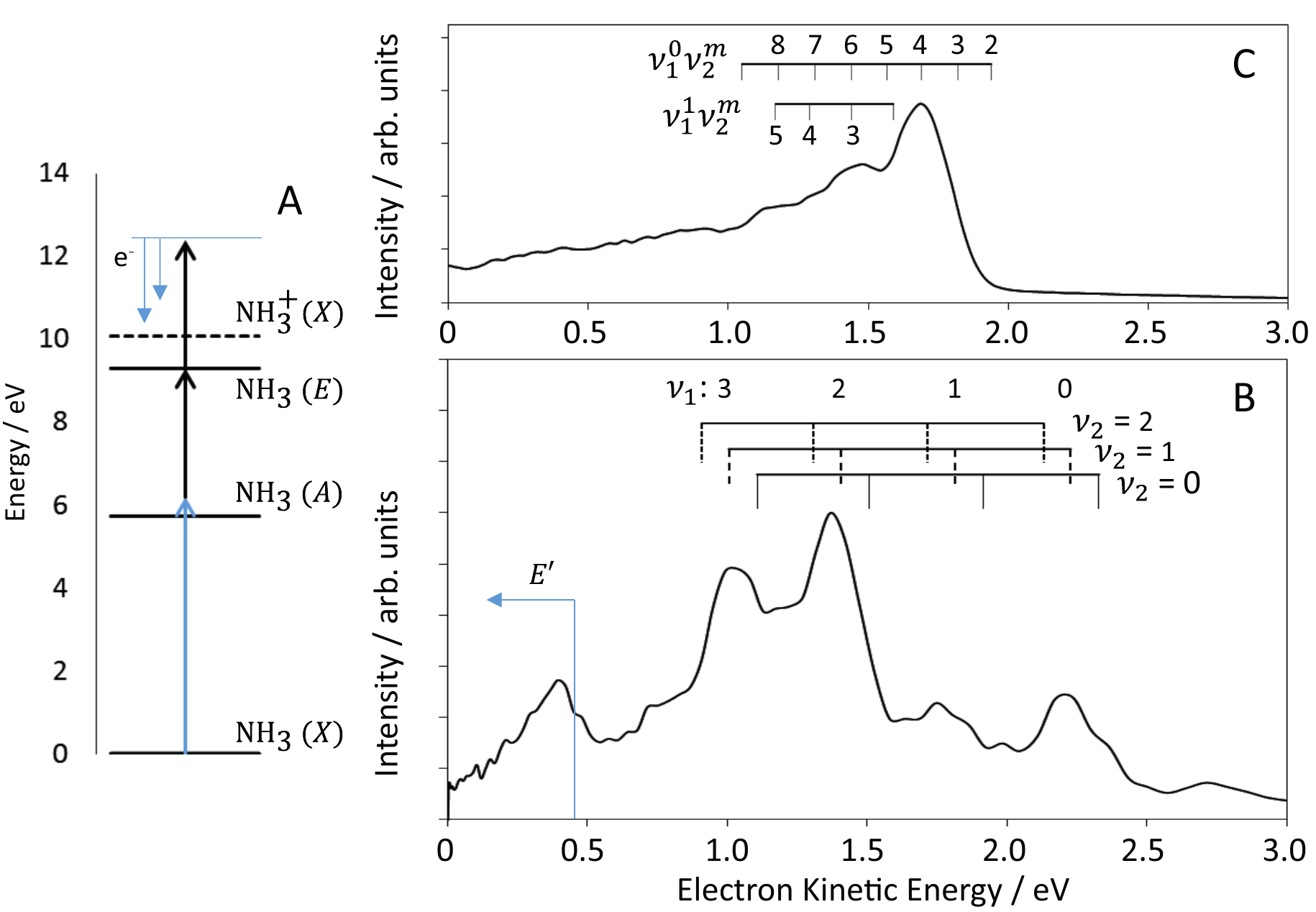}
\caption{A. Excitation scheme showing the resonant absorption processes and the ground ion state. B. 1+2$'$ Photoelectron spectrum and assignment of the time zero photoelectron spectrum. The combs above the plot show the dominant vibrational states in the photoelectron spectrum originating from the ground ion state. The region below 0.4 eV is assigned to the 2E$'$ excited state of the ion. C. 200.8 nm 1+1 Photoelectron spectrum and assignment as a comparison to the 1+2$'$ photoelectron spectrum in B. }
\label{PES-ass}
\end{figure}

In figure \ref{PES-ass} B we plot the background subtracted photoelectron spectrum for NH$_{3}$ obtained when the pump and probe are time overlapped. The pump populates the $\nu_2 ^4$ vibrational level of the $\tilde{A}$-state, which is subsequently ionised by absorption of 2 or 3 400 nm photons. Absorption of two photons leads to a total photon energy of 12.47 eV which is 2.4 eV above the ionisation potential for the ground, 2A$'$, ion state. Absorption of a third photon increases the available energy to 15.6 eV which is 0.39 eV above the vertical excitation energy of the 2E$'$ ion state providing another possible ionisation channel. The spectrum therefore contains features corresponding to two ion states and shows clear vibrational structure originating from the electronic ground state of the ion. The intensity of the probe is kept to a level that minimises two photon non-resonant excitation of the $\tilde{A}$-state. This means we do not observe any dynamics at negative time delays and means any probe-pump contributions to the spectrum are minimal. This is confirmed by the fact we obtain the same, albeit noisier, photoelectron spectrum at times outside the cross-correlation.

At high electron kinetic energies (< 0.5 eV) a clear vibrational progression is observed corresponding to ionisation into vibrationally excited states of the 2A$'$ ground ion state. We assign the vibrational features based on an anharmonic oscillator model of the form,

\begin{equation}
%\begin{split}
\label{aharm}
E = \omega_{e1} (\textit{v}_1 + \frac{1}{2}) - \omega_{e1} \chi_{e1} (\textit{v}_1 + \frac{1}{2})^2 \\
 + \omega_{e2} (\textit{v}_2 + \frac{1}{2}) - \omega_{e2} \chi_{e2} (\textit{v}_2 + \frac{1}{2})^2
%\end{split}
\end{equation}

where $\omega_{ei}$ and $\omega_{ei} \chi_{ei}$ are the harmonic and anharmonic vibrational frequencies of the ith vibrational mode of the ground ion state of ammonia and $\textit{v}_{i}$ is the vibrational quantum number of the state populated. The literature values used are taken from Xie \textit{et al}\cite{XieFD2000} and result in the expected kinetic energies given in table \ref{energy} and assignments shown in figure \ref{PES-ass} as the combs above the data. The assignments highlight a progression in the symmetric stretch, $\nu_1$, in combination with various quanta in the umbrella mode, $\nu_2$. Based on the calculated positions, the $\nu_2$=1 peaks show the highest intensity at all levels of excitation of $\nu_1$ with the maximum seen at the $\nu_1^2$$\nu_2^1$ level. The strongly overlapping nature of the features means that this is a rough assignment and slightly shifts the measured peak maxima from what may be expected for well resolved peaks. As such we define each of the features by the level of excitation in the symmetric stretch only. Each feature in the spectrum associated with the 2A$'$ state is therefore treated as having a different quanta of vibrational energy in $\nu_1$ and is labeled by this quantum number in the remainder of the paper.
The broad low energy feature peaking at 0.38 eV does not correlate well with the expected vibrational progression but correlates energetically with what would be expected following three photon ionisation into the 2E$'$ state. We therefore assign this feature to ionisation into the excited 2E$'$ state. The broad feature contains no vibrational structure and we therefore treat this as a single feature in the fits described later. There is also a small feature between 2.5 and 3 eV which is due to ionisation of the ammonia dimer which makes up a small component of our molecular beam.

\begin{table}[h]
\small
  \caption{\ Expected electron kinetic energies in eV for the various vibrational states in the NH$_3^+$ state calculated using equation \ref{aharm}. $\nu_1$ labels the symmetric stretch and $\nu_2$ labels the umbrella vibration.}
  \label{energy}
  \begin{tabular*}{0.5\textwidth}{@{\extracolsep{\fill}}llll}
    \hline
     & $\nu_1$ = 0 &  $\nu_1$ = 1 &  $\nu_1$ = 2 \\
    \hline
    $\nu_2$ = 0 & 2.33 & 2.23 & 2.13 \\
    $\nu_2$ = 1 & 1.92 & 1.81  & 1.72 \\
    $\nu_2$ = 2 & 1.51 & 1.41 & 1.31 \\
    $\nu_2$ = 3 & 1.11 & 1.01 & 0.91 \\
        \hline
  \end{tabular*}
\end{table}

\begin{figure} \centering
\includegraphics[width=0.7\columnwidth]{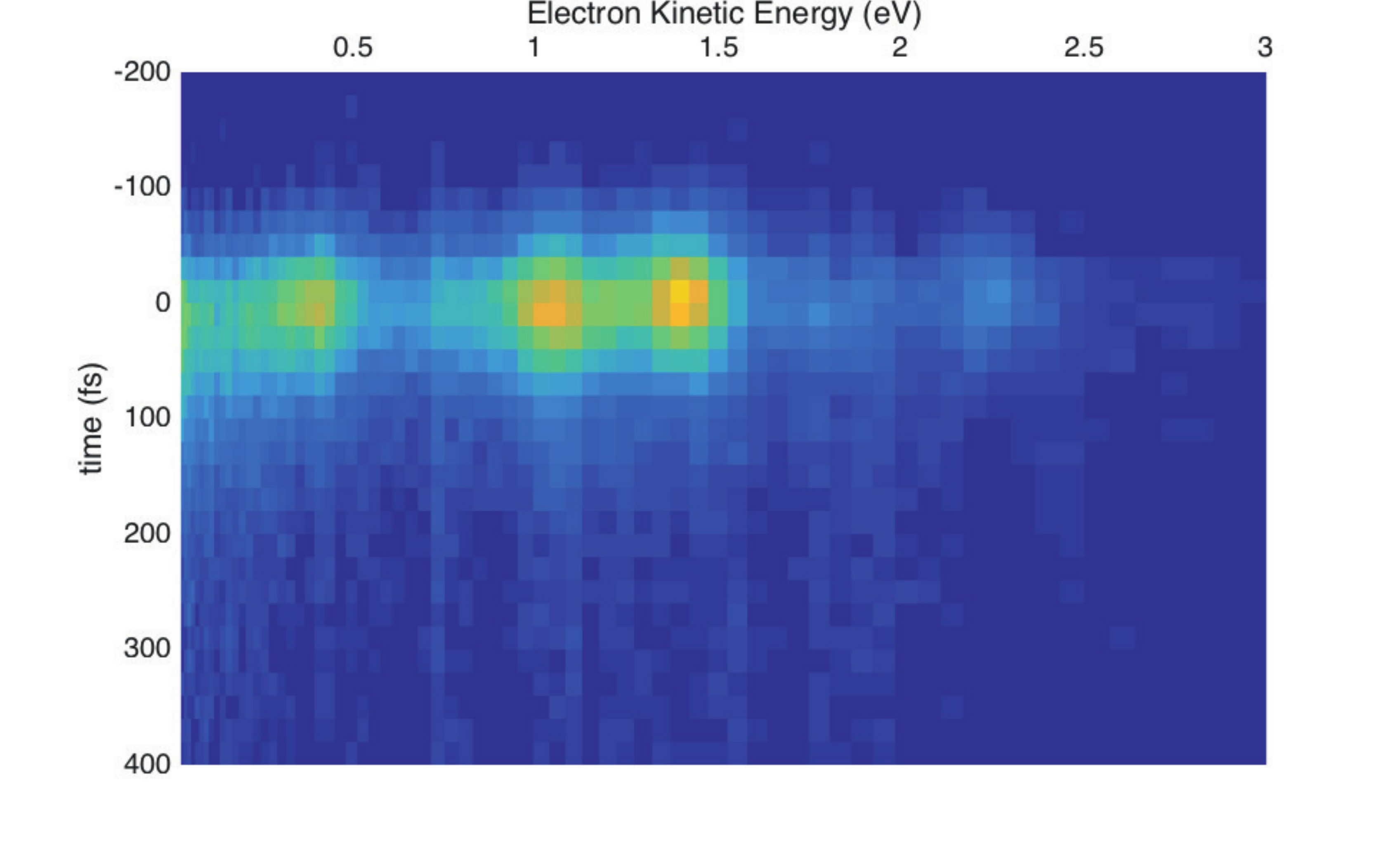}
\caption{colourmap of the time and energy dependent photoelectron spectrum obtained following excitation of the $\nu_2$=4 vibrational level of the electronic A-state and ionisation with multiple 400 nm photons.}
\label{PES-surf}
\end{figure}

\begin{figure} \centering
 \includegraphics[width=0.35\columnwidth]{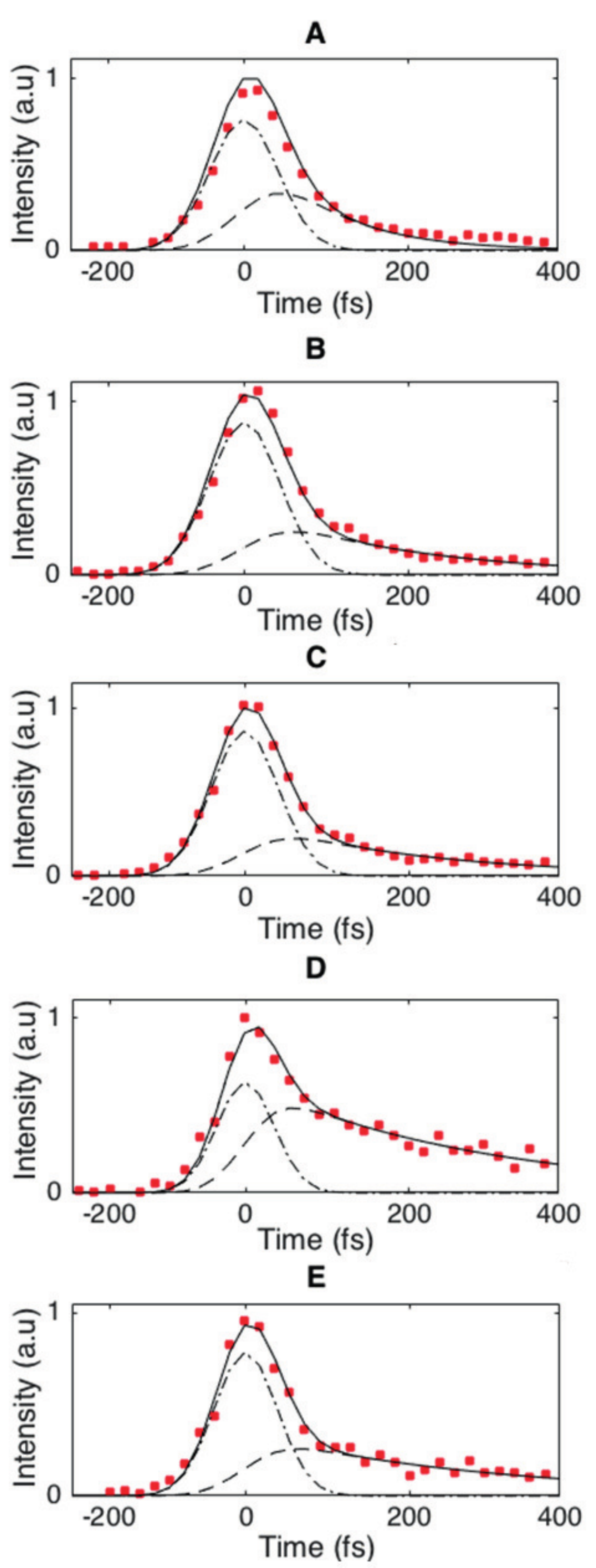}
 \caption{Experimentally measured (data points) time dependence of the photoelectron intensity for the main features observed in the photoelectron spectrum as well as the corresponding fits to equation \ref{fits} (solid line) and the Gaussian IRF (dot dash line) and exponentially decaying (dashed line) components of the fit. The plots correspond to the ionisation into the (A) 2E$'$, (B) 2A$'$ $\nu_1^3$, (C) 2A$'$$\nu_1^2$, (D) 2A$'$ $\nu_1^1$, (E) 2A$'$ $\nu_1^0$ ion states. The energy regions used and lifetimes obtained are given in table \ref{lifetime}}
 \label{exp1}
 \end{figure}

By comparing the photoelectron spectrum obtained with single photon ionisation measurements of the $\tilde{A}$-state we see very clear differences. In figure \ref{PES-ass} C we plot the 1+1 photoelectron spectrum at 200.8 nm. This provides almost exactly the same energy as the 1+2$'$ ionisation process and spectrum shown in figure \ref{PES-ass} B but has a starkly different apppearance. In the direct ionisation measurements, a strong $\Delta\textit{v}$=0 propensity is observed, such that ionisation from the $\nu_2^4$ vibrational level predominantly leads to population of the $\nu_2^4$ level in the ion with lower energy vibrational levels having negligible intensity in the observed photoelectron spectrum.\cite{Stavros2012} The vibrational progression in the single photon experiments is in the $\nu_2$ mode with at most one qaunta in the $\nu_1$ mode, such that the spectrum is dominated by the $\nu_1^0$$\nu_2^m$ progression.\cite{Stavros2012} This is contrary to our findings where the dominant vibrational progression correlates with a change in the vibrational quantum number associated with $\nu_1$ with high levels of vibrational excitation in this mode. This is in conjunction with a progression in $\nu_2$ with a dominant transition into the $\nu_2$=1 level for all levels of $\nu_1$. The reason for the dramatic changes is the resonant excitation process at the single photon level. Absorption of a single 400 nm photon excites the $E'$ $^1$A$_1'$ Rydberg state. The resonant excitation and finite duration of the laser pulses changes the FC factors associated with ionisation, enhancing overlap with a different set of vibrational states. The states observed broadly match those presented in \cite{Lou2006} where the $E'$ Rydberg state is accessed through a non-resonant two photon process at 266 nm, followed by ionisation with a time delayed 400 nm photon. The spectrum is however not identical to that obtained from directly accessing the $E'$ Rydberg state, but contains a fingerprint associated with the $\tilde{A}$-state character. The similarity of the spectra does however confirm that ionisation proceeds through the intermediate $E'$ Rydberg state. The relatively complex spectrum obtained from ionisation of the Rydberg state is somewhat contrary to what may be expected. Ionisation of Rydberg states is often associated with strong $\Delta\textit{v}$=0 propensity due to the strong similarity of character between the Rydberg and ion state. This often leads to a strong single feature in the photoelectron spectrum obtained. The $E'$ state shows a more complex propensity with previous measurements showing the dominant transition is of $\Delta\textit{v}$=1 character, suggesting there are significant differences in the shape of the $E'$ and ion states.\cite{Lou2006} The difference in ionisation propensity for the $\tilde{A}$ and $E'$ states means the transition between them is likely to involve a change in vibrational state and is the reason for the rather different appearance of the two photoelectron spectra.

In figure \ref{PES-surf} we plot the time dependence of the full photoelectron spectrum. Each peak in the photoelectron spectrum rises together followed by what appears to be a biexponetial decay. In common with previous single photon ionisation experiments we do not observe any energy shifts as a function of time.\cite{Yu2014,Stavros2012} To highlight the decay dynamics we plot lineouts for each of the features associated with the various levels of $\nu_1$ excitation. In each case the early time signal follows the intensity profile of the instrument response function (IRF) which is obtained from the cross-correlation of our pump and probe pulses. This corresponds to a Gaussian of full width at half maximum (FWHM) of 108 fs. In the longer term, the dynamics show an exponential decay such that we fit the overall experimental signal as a combination of a Gaussian, corresponding to our IRF, plus a component which is the Gaussian IRF convoluted with an exponential decay,

\begin{equation}
 \label{fits}
 I(t) =  N_2 G(\sigma t) +N_1 G(\sigma t) \ast e^{\lambda t}
 \end{equation}

Where G is the Gaussian IRF with a FWHM of $\sigma$, $t$ is the pump probe delay and $\lambda$ is the fitted decay constant. $N_1$ and $N_2$ are normalisation constants of the two signal components. The Gaussian contribution to the overall signal is due to strong field ionisation as the two laser pulses overlap in time and provides a fingerprint for ionisation from the FC region. Information about the dynamics is then contained within the lifetimes obtained from the second component in the fit. Each feature was fitted simultaneously using a least squares method, with the fits to equation \ref{fits} plotted in figure \ref{exp1} along with the Gaussian IRF and the Gaussian convoluted with an exponential decay. The energy ranges used and lifetimes (1/$\lambda$) obtained from the fits are also given in table \ref{lifetime}. The plots and table show that each of the vibrational features has a different associated lifetime suggesting that the FC factors associated with ionisation are changing as a function of time and that these could provide a sensitive probe of the excited state dynamics.

\begin{table}[h]
\small
  \caption{\ Decay lifetimes derived from the fits of the experimental data to equation \ref{fits}.}
  \label{lifetime}
  \begin{tabular*}{0.5\textwidth}{@{\extracolsep{\fill}}llll}
    \hline
    Energy range / eV & Ion state &  Vibrational state &  Lifetime / fs \\
    \hline
    0.34-0.4 & 2E$'$ &  - & 184 $\pm$ 14\\
    1.0-1.06 & 2A$'$ & $\nu_1^3$  & 212 $\pm$ 17\\
    1.32-1.5 & 2A$'$ & $\nu_1^2$ & 228 $\pm$ 25\\
    1.69-1.81 & 2A$'$ & $\nu_1^1$ & 307 $\pm$ 26\\
    2.26-2.44 & 2A$'$ & $\nu_1^0$ & 325 $\pm$ 54\\
    \hline
  \end{tabular*}
\end{table}

The fits provide a different lifetime for each of the features associated with each electronic and vibrational state of the ion. Longer lifetimes are found for features of higher electron kinetic energy correlating with lower levels of excitation in the symmetric stretch in the ground state of the ion. The shortest lifetime is associated with ionisation into the 2E$'$ excited ion state at an electron kinetic energy of 0.3 eV which has a lifetime of 184 fs. The vibrational features associated with ionisation into the 2A$'$ state have a longer lifetime, with the lifetime seen to increase as the vibrational energy in the ion state is reduced. The highest internal energy feature of the 2A$'$ state, associated with the $\nu_1^3$ vibrational level, has a lifetime of 212 fs, while the lower energy $\nu_1^0$ has a lifetime of 325 fs. The lifetimes are significantly longer than the < 75 fs derived from photoelectron spectroscopy measurements using direct ionisation from the $\tilde{A}$-state.\cite{Stavros2012} A key point from all of these measurements is that the timescales derived correspond to the time it takes for the excited state wavepacket to leave the observation window of the experiment as usually set by the probe energy. The < 75 fs timescale derived previously has defined the time taken for the wavepacket to lose overlap with the accessible ion states and leave the initially populated FC region.\cite{Stavros2012} The longer timescales provided by this experiment more closely match those obtained in recent time and energy resolved ion yield measurements on the same vibrational state of ammonia.\cite{chatterley2013}\cite{Wells2009}  In the ion kinetic energy release work the appearance times of the hydrogen atom is resolved into various kinetic energy regions correlating with adiabatic and non-adiabatic dissociation processes. The kinetic model proposed broadly separated the dissociation trajectories in two; one centred on more planar geometries at the FC region which lead to non-adiabatic dissociation and the formation of ground state products; a second set of trajectories coming from non-planar geometries leading to formation of adiabatic dissociation products. Both sets of trajectories left the FC region and moved towards the conical intersection located at extended H-NH$_2$ bond lengths on a timescale in the region of 100 fs. Once in the region of the conical intersection, the trajectories originating from planar geometries were seen to relax through the conical intersection within 70 fs, leading to the rapid formation of ground state products through a non-adiabatic process. Conversely, the trajectories originating from non-planar geometries were seen to have an extended lifetime in the region of the conical intersection leading to a delayed appearance of the adiabatic dissociation products. The delayed appearance allowed for a lifetime of the population in the region of the conical intersection to be derived which was on the order of 400 fs.\cite{chatterley2013}

\begin{figure} \centering
 \includegraphics[width=0.5\columnwidth]{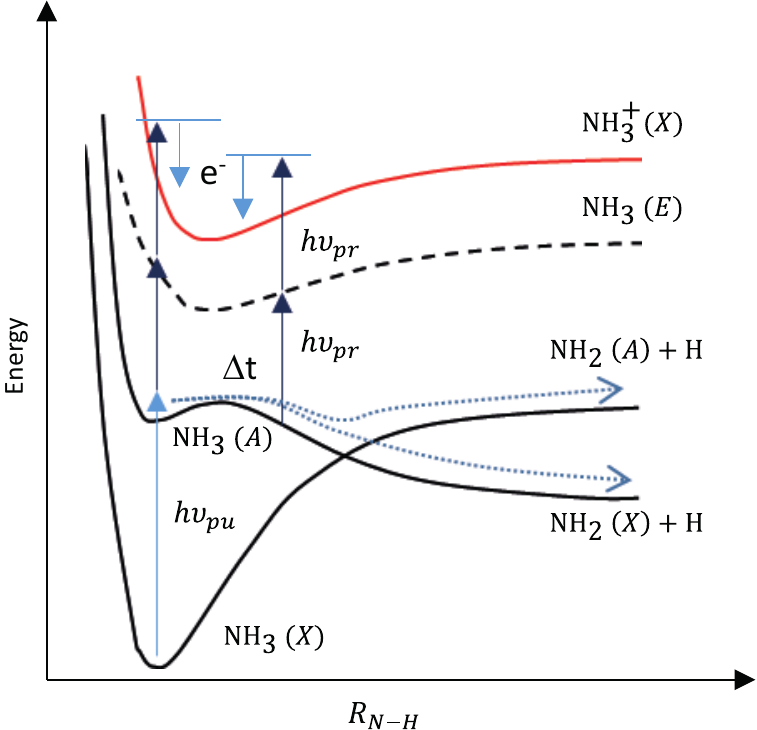}
 \caption{Schematic representation of the potentials of ammonia of relevence to the experiment. A single UV photon at 200 nm, $h\nu_{pu}$, excites the $\nu_2$=4 vibrational level of the $\tilde{A}$-state. Absorpion of a single 400 nm photon, $h\nu_{pr}$ populates vibrational levels of the $E'$ Rydberg state. At early times the dominant transition into the $E'$ Rydberg state leads to population of the $\nu_1$=2 vibrational level of the ion state upon absorption of a second 400 nm photon. After approximately 100 fs the wavepacket is localised in the region of the conical intersection where different trajectories to the adiabatic and non-adiabatic dissociation products give rise to different levels of vibrational excitation in the $E'$ Rydberg state. These manifest as changes in the vibrational lifetimes of the measured ion states accessed following the absorption of a second 400 nm photon.}
 \label{potentials}
 \end{figure}

The timescales derived from the photoion kinetic energy release measurements are therefore at odds with previous photoelectron spectroscopy measurements and slightly longer than those measured in our experiment. As mentioned above, the difference lies in the measurements process and what part of the full dynamics is being measured. Based on the timescales the resonant multiphoton probe is sensitive to an extended region of the excited state potential energy surface but is not sensitive to the final dissociation product formation. The difference in lifetimes obtained for the various vibrational states are a consequence of the resonant ionisation process. Rydberg states are in general very similar in shape to the ion states to which they converge. Rydberg states therefore have a very strong $\Delta\textit{v}$=0 transition propensity which gives rise to extremely sharp lines in their photoelectron spectra. This is a key property in the use of Rydberg fingerprint spectroscopy experiments where often the use of superexcited and Rydberg states provides detailed information about static and dynamic structure.\cite{Weber2015,Weber2005} The situation in the $E'$ state is slightly more complex, however the changes in the observed photoelectron spectrum can still be related back to the dynamics in the $\tilde{A}$-state. The change in vibrational levels seen in the time dependent photoelectron spectrum is related to which vibrational levels of the Rydberg state are accessed after the absorption of a single probe photon and can be rationalised as follows and with reference to figure \ref{potentials}. At early times, absorption of a single 400 nm photon populates a particular set of vibrational levels in the $E'$  Rydberg state. The dominant level populated leads to the photoelectron spectrum obtained at early times, peaking at the $\nu_1^2$$\nu_2^1$ level. As the wavepacket propagates on the $\tilde{A}$-state towards the conical intersection and dissociation, the various trajectories lead to changes in the dominant transition into the $E'$ Rydberg state. Changes in the transition probability into the Rydberg state then manifest in the photoelectron spectrum.

The higher vibrational levels associated with the symmetric stretch show a shorter lifetime. We associate this lifetime with the non-adiabatic dissociation process which, from our measurements, provides an average lifetime around 200 fs. This closely matches the average lifetimes obtained from ion kinetic energy release measurements which showed a lifetime between 82 fs and 263 fs. The different lifetimes were associated with different levels of vibrational excitation in the NH$_2$ fragment with the majority of the signal seen at higher levels of vibrational excitation correlating with the longer lifetime. The ~200 fs lifetime obtained from our photoelectron spectroscopy measurements would provide an average of these numbers and therefore correlates well with the excited state lifetime suggested from ion kinetic energy release measurements. With the current data we cannot correlate the vibrational levels in the photoelectron spectrum with the associated levels of vibrational excitation in the resulting NH$_2$ fragment. This may be possible in a more resolved spectrum or if calculations of the intermediate Rydberg state can provide insight into the resonance condition for the excitation and ionisation process. The non-adiabatic dissociation process is thought to predominantly originate from planar geometries whose trajectories we suggest maintain overlap with higher vibrational levels in the $E'$ Rydberg state out to the region of the conical intersection.

The lower vibrational states observed in the photoelectron spectrum have a much longer lifetime of over 300 fs. This is longer than the lifetime associated with non-adiabatic dissociation and more closely matches the lifetime associated with adiabatic dissociation into an electronically excited NH$_2$ fragment in conjunction with H. The adiabatic dissociation products are considered to be formed by trajectories originating from non-planar geometries in the $\tilde{A}$-state. These maybe expected to follow a less direct trajectory towards the conical intersection and dissociation such that lower levels of the stretching vibration are observed. The 300 fs lifetime is someway short of the lifetime retrieved from ion kinetic energy release measurements, however, the photoelectron spectroscopy experiments will only be sensitive to regions where the probe photon remains resonant. While this will extend some distance along the excited state potential it is unlikely to extend beyond the region of the conical intersection. As the trajectory may be less direct this could explain the relatively short lifetime obtained when compared with the time at which the fully dissociated products are formed.

\section{Conclusions}
The photodissociation dynamics of ammonia in the $\nu_2^4$ vibrational level of the $\tilde{A}$-state have been studied using a resonant multiphoton ionisation probe in a photoelectron spectroscopy experiment. By probing the dynamics via the intermediate $E'$ Rydberg state we access a different set of vibrational levels in the ion and maintain overlap with the ion state for an extended period of time. By maintaining the overlap with the ion state without requiring a high energy probe, the observation window is maximised without the problems associated with DUV or XUV probes, maintaining a low background signal and clear early time dynamics. The dynamics in the $\tilde{A}$-state manifest as changes in the vibrational transition into the Rydberg state which show up in the measurement as vibrational state dependent lifetimes. Higher lying ion vibrational states have an average excited state lifetime of approximately 200 fs which closely matches the lifetime associated with the non-adiabatic dissociation process while the lower vibrational states have a much longer lifetime of over 300 fs which more closely matches the adiabatic dissociation process. The different trajectories taken by the two competing dissociation pathways at the region of the conical intersection show up different excitation and ionisation propensity allowing us to isolate features related to each process.

\section*{Acknowledgments}
All authors thank the STFC for access to the Artemis facility (app. number 13220015). RSM thanks the Royal Society for a University Research Fellowship (UF100047) and the Leverhulme trust for research support and for ADS's studentship (RPG-2013-365). HMW thanks the Central Laser Facility and Chemistry at the University of Southampton for a studentship. EJ thanks Chemistry at the University of Southampton for a studentship. We also acknowledge funding from the EC's Seventh Framework Programme (LASERLAB-EUROPE, grant agreement n$^\circ$ 228334). We thank Phil Rice for technical assistance.

\bibliography{NH3PCCP}

\providecommand*{\mcitethebibliography}{\thebibliography}
\csname @ifundefined\endcsname{endmcitethebibliography}
{\let\endmcitethebibliography\endthebibliography}{}
\begin{mcitethebibliography}{33}
\providecommand*{\natexlab}[1]{#1}
\providecommand*{\mciteSetBstSublistMode}[1]{}
\providecommand*{\mciteSetBstMaxWidthForm}[2]{}
\providecommand*{\mciteBstWouldAddEndPuncttrue}
  {\def\EndOfBibitem{\unskip.}}
\providecommand*{\mciteBstWouldAddEndPunctfalse}
  {\let\EndOfBibitem\relax}
\providecommand*{\mciteSetBstMidEndSepPunct}[3]{}
\providecommand*{\mciteSetBstSublistLabelBeginEnd}[3]{}
\providecommand*{\EndOfBibitem}{}
\mciteSetBstSublistMode{f}
\mciteSetBstMaxWidthForm{subitem}
{(\emph{\alph{mcitesubitemcount}})}
\mciteSetBstSublistLabelBeginEnd{\mcitemaxwidthsubitemform\space}
{\relax}{\relax}

\bibitem[Evans \emph{et~al.}({2012})Evans, Yu, Roberts, Stavros, and
  Ullrich]{Stavros2012}
N.~L. Evans, H.~Yu, G.~M. Roberts, V.~G. Stavros and S.~Ullrich,
  \emph{{Physical Chemistry Chemical Physics}}, {2012}, \textbf{{14}},
  {10401--10409}\relax
\mciteBstWouldAddEndPuncttrue
\mciteSetBstMidEndSepPunct{\mcitedefaultmidpunct}
{\mcitedefaultendpunct}{\mcitedefaultseppunct}\relax
\EndOfBibitem
\bibitem[Bonhommeau \emph{et~al.}({2009})Bonhommeau, Valero, Truhlar, and
  Jasper]{Truhlar2009}
D.~Bonhommeau, R.~Valero, D.~G. Truhlar and A.~W. Jasper, \emph{{Journal of
  Chemical Physics}}, {2009}, \textbf{{130}}, {234303}\relax
\mciteBstWouldAddEndPuncttrue
\mciteSetBstMidEndSepPunct{\mcitedefaultmidpunct}
{\mcitedefaultendpunct}{\mcitedefaultseppunct}\relax
\EndOfBibitem
\bibitem[Wells \emph{et~al.}({2009})Wells, Perriam, and Stavros]{Wells2009}
K.~L. Wells, G.~Perriam and V.~G. Stavros, \emph{{Journal of Chemical
  Physics}}, {2009}, \textbf{{130}}, {074308}\relax
\mciteBstWouldAddEndPuncttrue
\mciteSetBstMidEndSepPunct{\mcitedefaultmidpunct}
{\mcitedefaultendpunct}{\mcitedefaultseppunct}\relax
\EndOfBibitem
\bibitem[Hause \emph{et~al.}({2008})Hause, Yoon, and Crim]{Crim2008}
M.~L. Hause, Y.~H. Yoon and F.~F. Crim, \emph{{Molecular Physics}}, {2008},
  \textbf{{106}}, {1127--1133}\relax
\mciteBstWouldAddEndPuncttrue
\mciteSetBstMidEndSepPunct{\mcitedefaultmidpunct}
{\mcitedefaultendpunct}{\mcitedefaultseppunct}\relax
\EndOfBibitem
\bibitem[Li \emph{et~al.}({2007})Li, Valero, and Truhlar]{Truhlar2007}
Z.~H. Li, R.~Valero and D.~G. Truhlar, \emph{{Theoretical Chemistry Accounts}},
  {2007}, \textbf{{118}}, {9--24}\relax
\mciteBstWouldAddEndPuncttrue
\mciteSetBstMidEndSepPunct{\mcitedefaultmidpunct}
{\mcitedefaultendpunct}{\mcitedefaultseppunct}\relax
\EndOfBibitem
\bibitem[Hause \emph{et~al.}({2006})Hause, Yoon, and Crim]{Crim2006}
M.~L. Hause, Y.~H. Yoon and F.~F. Crim, \emph{{Journal of Chemical Physics}},
  {2006}, \textbf{{125}}, {174309}\relax
\mciteBstWouldAddEndPuncttrue
\mciteSetBstMidEndSepPunct{\mcitedefaultmidpunct}
{\mcitedefaultendpunct}{\mcitedefaultseppunct}\relax
\EndOfBibitem
\bibitem[Loomis \emph{et~al.}({2000})Loomis, Reid, and Leone]{Leone2000}
R.~Loomis, J.~Reid and S.~Leone, \emph{{Journal of Chemical Physics}}, {2000},
  \textbf{{112}}, {658--669}\relax
\mciteBstWouldAddEndPuncttrue
\mciteSetBstMidEndSepPunct{\mcitedefaultmidpunct}
{\mcitedefaultendpunct}{\mcitedefaultseppunct}\relax
\EndOfBibitem
\bibitem[Mordaunt \emph{et~al.}({1998})Mordaunt, Ashfold, and
  Dixon]{Mordaunt1998}
D.~Mordaunt, M.~Ashfold and R.~Dixon, \emph{{Journal of Chemical Physics}},
  {1998}, \textbf{{109}}, {7659--7662}\relax
\mciteBstWouldAddEndPuncttrue
\mciteSetBstMidEndSepPunct{\mcitedefaultmidpunct}
{\mcitedefaultendpunct}{\mcitedefaultseppunct}\relax
\EndOfBibitem
\bibitem[Biesner \emph{et~al.}({1989})Biesner, Schnieder, Ahlers, Xie, Welge,
  Ashfold, and Dixon]{Biesner89_2}
J.~Biesner, L.~Schnieder, G.~Ahlers, X.~Xie, K.~Welge, M.~Ashfold and R.~Dixon,
  \emph{{Journal of Chemical Physics}}, {1989}, \textbf{{91}},
  {2901--2911}\relax
\mciteBstWouldAddEndPuncttrue
\mciteSetBstMidEndSepPunct{\mcitedefaultmidpunct}
{\mcitedefaultendpunct}{\mcitedefaultseppunct}\relax
\EndOfBibitem
\bibitem[Ashfold \emph{et~al.}({1985})Ashfold, Bennett, and Dixon]{Ashfold85}
M.~Ashfold, C.~Bennett and R.~Dixon, \emph{{Chemical Physics}}, {1985},
  \textbf{{93}}, {293--306}\relax
\mciteBstWouldAddEndPuncttrue
\mciteSetBstMidEndSepPunct{\mcitedefaultmidpunct}
{\mcitedefaultendpunct}{\mcitedefaultseppunct}\relax
\EndOfBibitem
\bibitem[Ashfold \emph{et~al.}({1986})Ashfold, Bennett, and Dixon]{Ashfold86}
M.~Ashfold, C.~Bennett and R.~Dixon, \emph{{Faraday Discussions}}, {1986},
  \textbf{{82}}, {163--175}\relax
\mciteBstWouldAddEndPuncttrue
\mciteSetBstMidEndSepPunct{\mcitedefaultmidpunct}
{\mcitedefaultendpunct}{\mcitedefaultseppunct}\relax
\EndOfBibitem
\bibitem[Bach \emph{et~al.}({2003})Bach, Hutchison, Holiday, and
  Crim]{Bach2003}
A.~Bach, J.~Hutchison, R.~Holiday and F.~Crim, \emph{{Journal of Physical
  Chemistry A}}, {2003}, \textbf{{107}}, {10490--10496}\relax
\mciteBstWouldAddEndPuncttrue
\mciteSetBstMidEndSepPunct{\mcitedefaultmidpunct}
{\mcitedefaultendpunct}{\mcitedefaultseppunct}\relax
\EndOfBibitem
\bibitem[Bach \emph{et~al.}({2002})Bach, Hutchison, Holiday, and
  Crim]{BachVibrational2002}
A.~Bach, J.~Hutchison, R.~Holiday and F.~Crim, \emph{{Journal of Chemical
  Physics}}, {2002}, \textbf{{116}}, {4955--4961}\relax
\mciteBstWouldAddEndPuncttrue
\mciteSetBstMidEndSepPunct{\mcitedefaultmidpunct}
{\mcitedefaultendpunct}{\mcitedefaultseppunct}\relax
\EndOfBibitem
\bibitem[Dixon({1996})]{Dixon1996}
R.~Dixon, \emph{{Molecular Physics}}, {1996}, \textbf{{88}}, {949--977}\relax
\mciteBstWouldAddEndPuncttrue
\mciteSetBstMidEndSepPunct{\mcitedefaultmidpunct}
{\mcitedefaultendpunct}{\mcitedefaultseppunct}\relax
\EndOfBibitem
\bibitem[Seideman({1995})]{Seideman1995}
T.~Seideman, \emph{{Journal of Chemical Physics}}, {1995}, \textbf{{103}},
  {10556--10565}\relax
\mciteBstWouldAddEndPuncttrue
\mciteSetBstMidEndSepPunct{\mcitedefaultmidpunct}
{\mcitedefaultendpunct}{\mcitedefaultseppunct}\relax
\EndOfBibitem
\bibitem[Ziegler({1985})]{Ziegler1985}
L.~Ziegler, \emph{{Journal of Chemical Physics}}, {1985}, \textbf{{82}},
  {664--669}\relax
\mciteBstWouldAddEndPuncttrue
\mciteSetBstMidEndSepPunct{\mcitedefaultmidpunct}
{\mcitedefaultendpunct}{\mcitedefaultseppunct}\relax
\EndOfBibitem
\bibitem[Rodriguez \emph{et~al.}(2014)Rodriguez, Gonzalez, Rubio-Lago, and
  Banares]{Rodriguez2014}
J.~D. Rodriguez, M.~G. Gonzalez, L.~Rubio-Lago and L.~Banares, \emph{Phys.
  Chem. Chem. Phys.}, 2014, \textbf{16}, 406--413\relax
\mciteBstWouldAddEndPuncttrue
\mciteSetBstMidEndSepPunct{\mcitedefaultmidpunct}
{\mcitedefaultendpunct}{\mcitedefaultseppunct}\relax
\EndOfBibitem
\bibitem[Yu \emph{et~al.}(2014)Yu, Evans, Chatterley, Roberts, Stavros, and
  Ullrich]{Yu2014}
H.~Yu, N.~L. Evans, A.~S. Chatterley, G.~M. Roberts, V.~G. Stavros and
  S.~Ullrich, \emph{The Journal of Physical Chemistry A}, 2014, \textbf{118},
  9438--9444\relax
\mciteBstWouldAddEndPuncttrue
\mciteSetBstMidEndSepPunct{\mcitedefaultmidpunct}
{\mcitedefaultendpunct}{\mcitedefaultseppunct}\relax
\EndOfBibitem
\bibitem[Giri \emph{et~al.}({2011})Giri, Chapman, Sanz, and Worth]{Worth2011}
K.~Giri, E.~Chapman, C.~S. Sanz and G.~Worth, \emph{{Journal of Chemical
  Physics}}, {2011}, \textbf{{135}}, {044311}\relax
\mciteBstWouldAddEndPuncttrue
\mciteSetBstMidEndSepPunct{\mcitedefaultmidpunct}
{\mcitedefaultendpunct}{\mcitedefaultseppunct}\relax
\EndOfBibitem
\bibitem[Bernardi \emph{et~al.}(1996)Bernardi, Olivucci, and
  Robb]{bernardi_potential_1996}
F.~Bernardi, M.~Olivucci and M.~A. Robb, \emph{Chem. Soc. Rev.}, 1996,
  \textbf{25}, 321--328\relax
\mciteBstWouldAddEndPuncttrue
\mciteSetBstMidEndSepPunct{\mcitedefaultmidpunct}
{\mcitedefaultendpunct}{\mcitedefaultseppunct}\relax
\EndOfBibitem
\bibitem[Donnelly \emph{et~al.}(1979)Donnelly, Baronavski, and
  McDonald]{DONNELLY1979}
V.~Donnelly, A.~Baronavski and J.~McDonald, \emph{Chemical Physics}, 1979,
  \textbf{43}, 271 -- 281\relax
\mciteBstWouldAddEndPuncttrue
\mciteSetBstMidEndSepPunct{\mcitedefaultmidpunct}
{\mcitedefaultendpunct}{\mcitedefaultseppunct}\relax
\EndOfBibitem
\bibitem[Biesner \emph{et~al.}({1988})Biesner, Schnieder, Schmeer, Ahlers, Xie,
  Welge, Ashfold, and Dixon]{Biesner1988}
J.~Biesner, L.~Schnieder, J.~Schmeer, G.~Ahlers, X.~Xie, K.~Welge, M.~Ashfold
  and R.~Dixon, \emph{{Journal of Chemical Physics}}, {1988}, \textbf{{88}},
  {3607--3616}\relax
\mciteBstWouldAddEndPuncttrue
\mciteSetBstMidEndSepPunct{\mcitedefaultmidpunct}
{\mcitedefaultendpunct}{\mcitedefaultseppunct}\relax
\EndOfBibitem
\bibitem[Chatterley \emph{et~al.}(2013)Chatterley, Roberts, and
  Stavros]{chatterley2013}
A.~S. Chatterley, G.~M. Roberts and V.~G. Stavros, \emph{The Journal of
  Chemical Physics}, 2013, \textbf{139}, 034318\relax
\mciteBstWouldAddEndPuncttrue
\mciteSetBstMidEndSepPunct{\mcitedefaultmidpunct}
{\mcitedefaultendpunct}{\mcitedefaultseppunct}\relax
\EndOfBibitem
\bibitem[Makida \emph{et~al.}(2014)Makida, Igarashi, Fujiwara, Sekikawa,
  Harabuchi, and Taketsugu]{sekikawa1}
A.~Makida, H.~Igarashi, T.~Fujiwara, T.~Sekikawa, Y.~Harabuchi and
  T.~Taketsugu, \emph{The Journal of Physical Chemistry Letters}, 2014,
  \textbf{5}, 1760--1765\relax
\mciteBstWouldAddEndPuncttrue
\mciteSetBstMidEndSepPunct{\mcitedefaultmidpunct}
{\mcitedefaultendpunct}{\mcitedefaultseppunct}\relax
\EndOfBibitem
\bibitem[Iikubo \emph{et~al.}(2015)Iikubo, Fujiwara, Sekikawa, Harabuchi,
  Satoh, Taketsugu, and Kayanuma]{sekikawa2}
R.~Iikubo, T.~Fujiwara, T.~Sekikawa, Y.~Harabuchi, S.~Satoh, T.~Taketsugu and
  Y.~Kayanuma, \emph{The Journal of Physical Chemistry Letters}, 2015,
  \textbf{6}, 2463--2468\relax
\mciteBstWouldAddEndPuncttrue
\mciteSetBstMidEndSepPunct{\mcitedefaultmidpunct}
{\mcitedefaultendpunct}{\mcitedefaultseppunct}\relax
\EndOfBibitem
\bibitem[Nugent-Glandorf \emph{et~al.}(2001)Nugent-Glandorf, Scheer, Samuels,
  Mulhisen, Grant, Yang, Bierbaum, and Leone]{leoneXUV}
L.~Nugent-Glandorf, M.~Scheer, D.~A. Samuels, A.~M. Mulhisen, E.~R. Grant,
  X.~Yang, V.~M. Bierbaum and S.~R. Leone, \emph{Phys. Rev. Lett.}, 2001,
  \textbf{87}, 193002\relax
\mciteBstWouldAddEndPuncttrue
\mciteSetBstMidEndSepPunct{\mcitedefaultmidpunct}
{\mcitedefaultendpunct}{\mcitedefaultseppunct}\relax
\EndOfBibitem
\bibitem[Irimia \emph{et~al.}({2009})Irimia, Dobrikov, Kortekaas, Voet, van~den
  Ende, Groen, and Janssen]{Janssen2009}
D.~Irimia, D.~Dobrikov, R.~Kortekaas, H.~Voet, D.~A. van~den Ende, W.~A. Groen
  and M.~H.~M. Janssen, \emph{{Review of Scientific Instruments}}, {2009},
  \textbf{{80}}, {113303}\relax
\mciteBstWouldAddEndPuncttrue
\mciteSetBstMidEndSepPunct{\mcitedefaultmidpunct}
{\mcitedefaultendpunct}{\mcitedefaultseppunct}\relax
\EndOfBibitem
\bibitem[Eppink and Parker({1997})]{Eppink1997}
A.~Eppink and D.~Parker, \emph{{Review of Scientific Instruments}}, {1997},
  \textbf{{68}}, {3477--3484}\relax
\mciteBstWouldAddEndPuncttrue
\mciteSetBstMidEndSepPunct{\mcitedefaultmidpunct}
{\mcitedefaultendpunct}{\mcitedefaultseppunct}\relax
\EndOfBibitem
\bibitem[Roberts \emph{et~al.}({2009})Roberts, Nixon, Lecointre, Wrede, and
  Verlet]{POP}
G.~M. Roberts, J.~L. Nixon, J.~Lecointre, E.~Wrede and J.~R.~R. Verlet,
  \emph{{Review of Scientific Instruments}}, {2009}, \textbf{{80}},
  {053104}\relax
\mciteBstWouldAddEndPuncttrue
\mciteSetBstMidEndSepPunct{\mcitedefaultmidpunct}
{\mcitedefaultendpunct}{\mcitedefaultseppunct}\relax
\EndOfBibitem
\bibitem[Xie \emph{et~al.}({2000})Xie, Jiang, Li, Yang, Xu, Sha, Xu, Lou, and
  Zhang]{XieFD2000}
J.~Xie, B.~Jiang, G.~Li, S.~Yang, J.~Xu, G.~Sha, D.~Xu, N.~Lou and C.~Zhang,
  \emph{{Faraday Discussions}}, {2000}, \textbf{{115}}, {127--136}\relax
\mciteBstWouldAddEndPuncttrue
\mciteSetBstMidEndSepPunct{\mcitedefaultmidpunct}
{\mcitedefaultendpunct}{\mcitedefaultseppunct}\relax
\EndOfBibitem
\bibitem[Liu \emph{et~al.}({2006})Liu, Yin, Zhang, Wang, Jiang, and
  Lou]{Lou2006}
H.~P. Liu, S.~H. Yin, J.~Y. Zhang, L.~Wang, B.~Jiang and N.~Q. Lou,
  \emph{{Physical Review A}}, {2006}, \textbf{{74}}, {053418}\relax
\mciteBstWouldAddEndPuncttrue
\mciteSetBstMidEndSepPunct{\mcitedefaultmidpunct}
{\mcitedefaultendpunct}{\mcitedefaultseppunct}\relax
\EndOfBibitem
\bibitem[Pemberton \emph{et~al.}({2015})Pemberton, Zhang, Saita, Kirrander, and
  Weber]{Weber2015}
C.~C. Pemberton, Y.~Zhang, K.~Saita, A.~Kirrander and P.~M. Weber,
  \emph{{Journal of Physical Chemistry A}}, {2015}, \textbf{{119}},
  {8832--8845}\relax
\mciteBstWouldAddEndPuncttrue
\mciteSetBstMidEndSepPunct{\mcitedefaultmidpunct}
{\mcitedefaultendpunct}{\mcitedefaultseppunct}\relax
\EndOfBibitem
\bibitem[Gosselin and Weber({2005})]{Weber2005}
J.~Gosselin and P.~Weber, \emph{{Journal of Physical Chemistry A}}, {2005},
  \textbf{{109}}, {4899--4904}\relax
\mciteBstWouldAddEndPuncttrue
\mciteSetBstMidEndSepPunct{\mcitedefaultmidpunct}
{\mcitedefaultendpunct}{\mcitedefaultseppunct}\relax
\EndOfBibitem
\end{mcitethebibliography}
\bibliographystyle{rsc}

\end{document}